\documentstyle[11pt]{article}
\begin{document}
\thispagestyle{empty}
\baselineskip 20pt
\rightline{CU-TP-877, {\tt hep-th}/9802012}
\rightline{February, 1998} 

\

\def\tr{{\rm tr}\,}
\newcommand{\beq}{\begin{equation}}
\newcommand{\eeq}{\end{equation}}
\newcommand{\beqn}{\begin{eqnarray}}
\newcommand{\eeqn}{\end{eqnarray}}
\newcommand{\bde}{{\bf e}}
\newcommand{\balpha}{{\mbox{\boldmath $\alpha$}}}
\newcommand{\bsalpha}{{\mbox{\boldmath $\scriptstyle\alpha$}}}
\newcommand{\bbeta}{{\mbox{\boldmath $\beta$}}}
\newcommand{\bsbeta}{{\mbox{\boldmath $\scriptstyle\beta$}}}
\newcommand{\blambda}{{\mbox{\boldmath $\lambda$}}}
\newcommand{\bslambda}{{\mbox{\boldmath $\scriptstyle\lambda$}}}
\newcommand{\ggg}{{\boldmath \gamma}}
\newcommand{\ddd}{{\boldmath \delta}}
\newcommand{\mmm}{{\boldmath \mu}}
\newcommand{\nnn}{{\boldmath \nu}}

\newcommand{\bra}[1]{\langle {#1}|}
\newcommand{\ket}[1]{|{#1}\rangle}
\newcommand{\sn}{{\rm sn}}
\newcommand{\cn}{{\rm cn}}
\newcommand{\dn}{{\rm dn}}
\newcommand{\diag}{{\rm diag}}

\

\vskip 1cm
\centerline{\Large\bf Instantons and Magnetic Monopoles on $R^3\times S^1$ } 
\centerline{\Large\bf  with Arbitrary Simple Gauge Groups}
\vskip 1.2cm
\centerline{\large\it
Kimyeong Lee\footnote{electronic mail: klee@phys.columbia.edu}}
\vskip 1mm
\centerline{Physics Department, Columbia University, New York, NY, 10027}
\vskip 1.5cm
\centerline{\bf ABSTRACT}
\vskip 2mm
\begin{quote}
{\baselineskip 16pt We investigate Yang-Mills theories with arbitrary
gauge group on $R^3\times S^1$, whose symmetry is spontaneously broken
by the Wilson loop.  We show that instantons are made of fundamental
magnetic monopoles, each of which has a corresponding root in the
extended Dynkin diagram. The number of constituent magnetic monopoles
for a single instanton is the dual Coxeter number of the gauge group,
which also accounts for the number of instanton zero modes.  In
addition, we show that there exists a novel type of the $S^1$
coordinate dependent magnetic monopole solutions in $G_2,F_4,E_8$.  }
\end{quote}

%\pacs{14.80.Hv,11.27.+d,14.40.-n}

\newpage

In recent years there has been some interest in the connection between
magnetic monopoles and instantons in theories with partially
compactified space\cite{piljin,murray,caloron}.  Especially by
employing the $T$-duality on $D$-branes, we have shown that a single
instanton in the theory with $SU_n$ gauge group on space $R^3\times
S^1$ can be interpreted as a composite of $n$ distinct fundamental
monopoles~\cite{piljin}. On $R^3$, there are $n-1$ fundamental
magnetic monopole solutions~\cite{erick}. On $R^3\times S^1$, there
exist an additional type of magnetic monopole solution associated with
the lowest negative root, which plays the key role in making the net
magnetic charge of the $n$ distinct monopoles to be zero.

In this note we generalize the above consideration to theories with
arbitrary simple gauge group $G$ of rank $r$. Our consideration here
is purely field theoretic. With nontrivial the Wilson loop
$P\exp(i\int dx_4 A_4)$ along $S^1$, one can break the gauge symmetry
maximally to abelian subgroups $U(1)^r$ with the rank $r$ of the gauge
group.  Similar to the $SU(N)$ case, we show that besides usual $r$
fundamental BPS monopoles~\cite{erick}, there exist an additional type
of fundamental BPS monopoles associated with the lowest negative root.
Especially for the $G_2, F_4, E_8$ groups, we find that these new monopoles
will have $x_4$-dependence which cannot be gauged away.  We then show
that a single instanton solution is a composite of these $r+1$
fundamental monopoles such that the net magnetic charge is zero.  The
set of constituent magnetic monopoles is unique and the number of the
total fundamental magnetic monopoles turns out to be the dual Coxeter
number, which is also consistent with the previously known zero mode
counting~\cite{bernard}.

Depending on how one views the compactified direction, different
physics emerges. When we regard $S^1$ as the Euclidean time direction,
physics is that of finite temperature Yang-Mills theories. Our
relation between instantons and magnetic monopoles may shed some light
on further understanding of chiral symmetry breaking and confinement
in QCD. (For a current effort in lattice gauge theory community, see,
for example, Ref.~\cite{lattice}.) . When we view $S^1$ as the
compactified space, our work may give further insights on the relation
in some exact results in supersymmetric Yang-Mills theories in four
and three dimensions~\cite{seiberg,cumrun}. When we view $R^3\times
S^1$ as space and assume that there is additional time direction, our
work has more direct connection with D-brane dynamics~\cite{piljin}.

On $R^3\times S^1$, the Euclidean path integral measure can be
restricted to the gauge fields which are single-valued with respect to
the coordinate $x_4$ of $S^1$. We choose the $x_4$ interval to be
$[0,2\pi]$ for convenience.  Allowed gauge transformations $U(x)$ can
be multivalued if the transformed gauge field $UA_\mu U^\dagger -i
\partial_\mu U U^\dagger$ remain single-valued. Such a large gauge
transformation is possible only if the gauge group has a nontrivial
center. Nontrivial Wilson loop arises as the fourth component of the
gauge field takes nonzero expectation value. We can choose the
direction of this to lie along the Cartan subalgebra $H_i (1\le i\le
r)$, 
\beq
<A_4> = {\bf h}\cdot {\bf H}.
\label{wilson}
\eeq
For any finite action configuration, we can require the gauge field to
approach the above asymptotic value at spatial infinity modulo
gauge. One can always choose a set of simple roots, $\bbeta_i (1\le
i\le r) $, so that
\beq
{\bf h}\cdot \bbeta_i \ge 0\,\,\,\,\,{\rm for\,\, any} \,\, i.
\label{chamber}
\eeq
This defines the so-called Weyl chamber in the root vector space.

In this paper consider the gauge symmetry is maximally broken to
$U(1)^r$ so that the inequality in Eq.~(\ref{chamber}) holds strictly. 
We consider all generators of the gauge group in the adjoint
representation.  We normalize the generators so that 
\beq
 \tr (H_i H_j) = c_2(G) \delta_{ij}\,\,\,\,{\rm and}\,\,\,
 \tr (E_\alpha^\dagger E_\beta) = c_2(G)\delta_{\alpha\beta}.
\eeq
The normalization factor $c_2(G)$ is  the quadratic Casimir
for the adjoint representation because 
\beqn
c_2(G) &=& tr (T^a)^2 \,\, ({\rm no\,\, sum}) \nonumber \\
&=&  \frac{1}{d(G)}\sum_{a=1}^{d(G)} \tr (T^a)^2 =
\sum_{a=1}^{d(G)} (T^a)^2.
\eeqn
There is still an overall normalization to be fixed. With the choice
that longest root vectors to have length one, $c_2(G)$ becomes {\it
the dual Coxeter number}, which is an positive integer~\cite{bernard}.

The Yang-Mills action on $R^3\times S^1$  has  a lower bound $8\pi
|m|/g^2$, where the  topological charge is
\beqn
\!\!m\!&=&\!\frac{1}{64\pi^2}\!\int d^4x\,
\epsilon_{\mu\nu\rho\sigma}F_{\mu\nu}^a 
F^a_{\rho\sigma}
\nonumber \\ 
&=& \!  \frac{1}{8\pi^2}\! \int_0^{2\pi}\!\!\!\!\!\! dx_4
\int_{S^2_\infty}\!\!\!\!\! \!\!dS_i
\biggl[
B_i^a A_4^a - \frac{1}{2}\epsilon_{ijk}A_j^a \dot{A}^a_k \biggr],
\label{topo}
\eeqn
with $B^a_i =
\frac{1}{2}\epsilon_{ijk}F_{ij}^a$. (Here we assume that the  gauge fields
do not have any singularities at interior region of space time.) When
$m>0$, the bound is saturated by the field configurations satisfying
the self-dual equations,
\beq
B_i^a = D_i A_4^a - \partial_4 A^a_i.
\label{selfdual}
\eeq
Magnetic monopole configurations with no $x_4$ dependence would have
the topological charge contribution from the usual BPS term of
Eq.~(\ref{topo}), and instantons which goes to the pure gauge at
infinity would have the the contribution from the last term of
Eq.~(\ref{topo}). In addition, the instanton topological charge would
be integer since it measures the homotopy of this gauge from
$S^1\times S^2_\infty$ to $G$.

On $R^4$, one can find an instanton solution of minimum topological
charge, $m=1$, by embedding the $SU(2)$ solution along one of the
longest root. The number of zero modes around a single instanton turns
out to be $4c_2(G)$,~\cite{bernard}. The number of zero modes
$4c_2(G)$ around a single instanton can be interpreted as a sum of
five modes for the position and scale and the rest for the global
gauge modes which changes the embedded solution. On $R^3\times S^1$,
one find a periodic instanton solution, which turns out to be a
special limit where the gauge symmetry is partially
restored~\cite{piljin, changhai, caloron}. We expect the number of
zero modes to be identical to the  $R^4$ case.

If we consider the self-dual configurations which are independent of
$x_4$, Eq.~(\ref{selfdual}) becomes the standard BPS equations for
magnetic monopoles with $A_4$ playing the role of the Higgs field.
For each root $\balpha$, there is a corresponding $SU(2)$ subgroup
\beqn
 & & t^1(\balpha) = \frac{1}{\sqrt{2\balpha^2}} (E_\alpha +
E_{-\alpha}),
\nonumber \\
& & t^2(\balpha) = \frac{-i}{\sqrt{2\balpha^2}}(E_\alpha -
E_{-\alpha}),
\nonumber \\
& & t^3(\balpha) = \balpha^* \cdot {\bf H},
\eeqn
where the dual of a root $\balpha$ is defined as
\beq
\balpha^* = \balpha/\balpha^2.
\eeq
With our normalization, $\balpha^*=\balpha$ for any long root.
We can embed the $SU(2)$ single monopole solution
along ${\bf t}(\balpha)$, 
\beqn
A_\mu(x) =& &  \biggl[{\bf h}- ( {\bf h}\cdot \balpha^*) \balpha
\biggr]\cdot {\bf H} \delta_{\mu 4} \nonumber \\
& & + t^a(\balpha) V^a_\mu({\bf r}; {\bf
h}\cdot \balpha), 
\label{time0}
\eeqn
where 
\beqn
& & V^a_4({\bf r};u) = \hat{r}^a \biggl[\frac{1}{r}-u\coth ur \biggr],
\nonumber \\
& & V_i^a({\bf r};u) = \epsilon_{aij}\hat{r}_j
\biggl[\frac{1}{r}-\frac{u}{\sinh ur} \biggr].  
\eeqn
The asymptotic behavior of this solution along the negative $x^3$ axis
is given as $A_4 \rightarrow ({\bf h} -\balpha^*/r)\cdot {\bf H} $ and
$B_i \rightarrow \delta_{i3} \balpha^*/r^2\cdot {\bf H}$, leading to
the topological charge $m = {\bf h}\cdot \balpha^*$.  Especially each
fundamental monopole corresponding to simple roots $\bbeta_i$ would
have topological charge
\beq
m_i \equiv {\bf h }\cdot \bbeta^*_i>0.
\eeq
It is well known that each fundamental monopole carries only four zero
modes among any fluctuations independent of $x_4$~\cite{erick}. Any
other BPS configurations would be composed with these monopoles. Of
course on $R^3\times S^1$, there can be additional zero modes in the
Kaluza-Klein modes~\cite{piljin}, more  about which we will discuss later.

{}From the experience with the $SU_n$ case~\cite{piljin}, we expect that
there exist $x_4$-dependent self-dual solutions which can be
identified as magnetic monopoles associated with the lowest negative
root $\bbeta_0$. This root together with simple roots $\bbeta_i$ can
be described by the extended Dynkin diagram.  Note that $\bbeta_0$ is
given uniquely by a linear combination of simple roots with positive
integer coefficients, and it is one of the longest roots.  In the 
$SU_n$ case, this BPS monopole solution for $\bbeta_0$ is 
equivalent to a $x_4$ independent solution under a large gauge
transformation.

In the $SU_n$ case the presence of this additional magnetic monopole
was crucial in showing that instantons are made of the $n$ distinct
magnetic monopoles so that net magnetic charge is zero. Here we assume
that it is possible to put these $n$ different self-dual solutions
together nonlinear way. Especially in $SU_2$ case, the $\bbeta_0$
monopole has opposite magnetic charge even though the configuration is
self-dual. The magnetic attractive force between $\bbeta_0$ and
$\bbeta_1$ monopole cancels the Higgs repulsive force between them.
In $SU_2$ case, we can construct the explicit field configuration for
a single instanton as a composite of $\bbeta_0$ and $\bbeta_1$
monopoles by using the Nahm's method~\cite{changhai}.

For a general simple gauge group, there are several questions to be
answered before one sees  that instantons are composed of magnetic
monopoles: (1) What is the $\bbeta_0$ magnetic monopole configuration?
(2) What is the maximum range of ${\bf h}$ in which all fundamental
magnetic monopoles carry only four zero modes? (3) What is the
constituent magnetic charge for a single instanton configuration? (4)
How do the topological charge and the number of zero modes work out?

Let us first give the answer to the second question, which we will
justify later. The range where each of  $r+1$ fundamental monopoles
have only four zero mode is given  by the inequality
\beq
{\bf h}\cdot \bbeta_i^* > 0 \,\,\,{\rm and}\;\;  {\bf h}\cdot (-
\bbeta_0^*) < 1.
\label{cell}
\eeq
We call this as {\it the fundamental cell}. Since $\bbeta_0$ is a
negative root, ${\bf h}\cdot (-\bbeta_0)>0$.  If the range of $x_4$
were $R$ rather than $2\pi$, then the last inequality would be ${\bf
h}\cdot(- \bbeta_0^*)< 2\pi/R$.

To find the $\bbeta_0$ monopole solutions, we start by considering the
group $SU(3)$. In this case, the simple roots are $\bbeta_1=
\bde_1-\bde_2$ and $\bbeta_2=\bde_2-\bde_3$ with $\bde_i$ such that
$\bde_i\cdot\bde_j = \delta_{ij}/2$. The lowest root is $\bbeta_0 =
-\bde_1+\bde_3$. Its center $Z_3$ is generated by the gauge
transformations
\beq
C_j = e^{4\pi i \lambda_j \cdot {\bf H} },
\eeq
where $\blambda_i$ are fundamental weights satisfying
\beq
2\blambda_i\cdot\bbeta^*_j = \delta_{ij}.
\eeq
In $SU_3$, $\lambda_1 = \frac{1}{3}(2\bde_1-\bde_2-\bde_3)$ and
$\blambda_2=\frac{1}{3}(\bde_1+\bde_2-2\bde_3)$.  (One can easily see
$C_j\, E_{\beta_k}\, C_j^\dagger = E_{\beta_k}$ for all $k$, implying
that $C_j$ belongs to the center.) On the fundamental representations,
for example, $C_1|\blambda_1> = e^{i\frac{4\pi}{3}}|\blambda_1>$ and
$C_1|\blambda_2> = e^{i\frac{2\pi}{3}}|\blambda_2>$. There are 
corresponding large gauge transformations,
\beq
U_j = e^{2ix_4 \lambda_j \cdot {\bf H}},
\label{global}
\eeq
such that $U_j(2\pi) = C_j$.  Under, for example, $U_1^\dagger$, the
asymptotic value ${\bf h}$ transforms to
\beq
{\bf h}_{\rm new} = {\bf h} - 2\blambda_1.
\label{hnew}
\eeq
For this new asymptotic value, we note that $ {\bf h}_{\rm new} \cdot
\bbeta_0^*= 1+ {\bf h}\cdot \bbeta_0^* > 0 $ and ${\bf h}_{\rm new}
\cdot \bbeta_2 = {\bf h}\cdot \bbeta_2 $ in the fundamental cell. 
On the other hand ${\bf h}_{\rm new}\cdot \bbeta_1<0$. Thus with
respect to ${\bf h}_{\rm new}$, $\bbeta_0$ and $\bbeta_2$ are simple
roots. One can then construct the $x_4$-independent $\bbeta_0$
monopole solution. This solution has the topological charge
\beq
m_0=1+{\bf h}\cdot \bbeta_0^*.
\label{m0}
\eeq
Gauge transforming back to the orginal ${\bf h}$ by $U_1$, we get the
time-dependent $\bbeta_0$ magnetic monopole solution,
\beqn
A_\mu(x;\bbeta_0) =& & \!\! [{\bf h}-({\bf h}_{new}\cdot
\bbeta_0^*)\bbeta_0] \cdot {\bf H} \delta_{\mu 4} \nonumber \\
& & + U_1\, t^a(\bbeta_0)
U_1^\dagger\,\, V_\mu ({\bf r}; m_0) ,
\label{time1}
\eeqn
where we used the fact $\bbeta_0^*=\bbeta_0$. Its topological charge
(\ref{topo}) is gauge invariant and so is still $m_0$.

This turns out to be a general feature for all groups with nontrivial
center. Table I shows the center of gauge groups. Clearly there is a
relation between the symmetry of the extended Dynkin diagram and the
center of the group. Except for $G_2,F_4,E_8$, there is always at
least one simple long root, say $\bbeta_1$, which symmetric to
$\bbeta_0$. The corresponding global gauge transformation $U_1$ in
Eq.~(\ref{global}) with the corresponding fundamental weight
$\blambda_1$ can be used to construct the $\bbeta_0$ solution and
given again by Eq.~(\ref{time1}). (Since we know all simple roots in
terms of orthogonal vectors\cite{lie}, we can check this explicitly
with  some labor.)

\begin{center}
\begin{small}
\begin{tabular}{ l|c}\hline
Group & Z(G) \\ \hline
$SU_n$, $n\ge 1$ & $Z_n$ \\
$SO_{2n+1}$, $n\ge 3$ & $Z_2$ \\
$SO_{2n}$, $n$ even, $n\ge 4$ & $Z_2 \oplus Z_2$ \\ 
$SO_{2n}$, $n$ odd, $n\ge 5$ & $Z_4$ \\
$Sp_{2n}$, $n\ge 2$ & $Z_2$ \\ 
$E_6$ & $Z_3$ \\ 
$E_7$ & $Z_2$ \\ 
$E_8, F_4, G_2$ & $\{1\}$ \\ \hline
\end{tabular}
\end{small} \\
\vspace{5mm}
{\bf Table I}: The center of the universal covering group $G$
\end{center}
\vspace{4mm}

For $G_2,F_4,E_8$, their center is trivial and the above method fails
to give the $\bbeta_0$ monopole solution.  The monopole solution for
$\bbeta_0$ for these groups should be genuinely $x_4$-dependent, which
cannot be gauged away.  By using the fact that $[2\blambda_1\cdot {\bf
H}, E_{\bsbeta_0}] = -E_{\bsbeta_0}$ in the $SU_3$ case, we rewrite the
$x_4$-dependent solution (\ref{time1}) as
\beqn
&& A_\mu(x;\bbeta_0) =  ({\bf h}-({\bf h}_{new}\cdot
\bbeta_0^*)\bbeta_0)\cdot {\bf H} \delta_{\mu 4} \nonumber \\
&&\;\;\; + e^{-ix_4
t^3(\beta_0)} \, t^a(\bbeta_0)  e^{ix_4 t^3(\beta_0)} 
 V_\mu ({\bf r}; {\bf h}\cdot \bbeta_0).
\label{time2}
\eeqn
Clearly the above solution is single-valued in $x_4$ since
$t^a(\bbeta_0)$ transforms with $t=1$ under itself. For all other  gauge
groups with nontrival center, the above argument works also.
Now we notice that the above solution is perfectly acceptable even in
$G_2,F_4,E_8$ cases as it is single-valued. Indeed along the negative
$x_3$ axis, the above spherically symmetric  solution has exactly
right asymptotics for the $\bbeta_0$ monopoles. 
Its topological charge in $G_2,F_4,E_8$ are still $m_0$
since the topological charge for this solution lies just in $SU(2)$
sector and its value is invariant under this $SU(2)$.

Having constructed explicit configurations for all $r+1$ fundamental
monopoles, we are now in the position to ask whether they have only
four zero modes. The zero mode analysis around spherically symmetric
magnetic monopoles solutions can be done by introducing spinors
$\psi=\delta A_4 +i \sigma_i \delta A_i$, which satisfy the Dirac
equation
\beq
(-i\sigma_j  D_j + D_4 ) \psi = 0.
\label{fluct}
\eeq
This is equivalent to the linear fluctuation equations of the
self-dual equation and the background gauge condition $D_\mu \delta
A_\mu=0$.  For the time-independent solution (\ref{time0}), the
analysis is quite similar to Ref.~\cite{erick}, except the
fluctuations can now depend on $x_4$, though single-valued, so that $\psi
\sim e^{ilx_4}$ with an integer $l$. The generators of the Lie algebra
can be split into irreducible representations of ${\bf
t}(\bbeta_i)$. The two relevant parameters along a root
$\alpha$~\cite{erick} are the isospin $t_3=\bbeta_i^*\cdot \balpha$
and the hypercharge
\beq
y = \frac{{\bf h}\cdot \balpha}{{\bf h}\cdot \bbeta_i} - t_3
+ \frac{l}{{\bf h}\cdot \bbeta_i}.
\label{ypara}
\eeq
The number of normalizable zero mode depends on the value of $y$ and
the isospin $t$ of the $\balpha$ multiplet with respect to ${\bf
t}(\bbeta_i)$. As given in Eq.(C.5) of Ref.~\cite{erick}, the $|y|$ ranges
for the nonzero number of zero modes for each $t$  are
\beqn 
& & t=\frac{1}{2};\,\, {\rm one \,\, for}\, |y|<\frac{1}{2}, \nonumber \\  
& & t=1;\, \,\,{\rm two \,\, for} \,\, |y|<1,  \nonumber \\
& & t=\frac{3}{2};\,\,{\rm four \,\, for} \,\, |y|<\frac{1}{2}, \,\,
{\rm three \,\, for}\,\, \frac{1}{2}< |y|< \frac{3}{2}. 
\eeqn
Otherwise with $t\le 3/2$ there is no zero mode. (At the boundary of
$|y|$ range, the matter is subtle but does not concern in this note.)
For example in $SU_3$, when ${\bf h}$ lies in the fundamental cell
(\ref{cell}), one can see easily that each of time-independent
$\bbeta_i$ monopoles has only four zero modes. Outside this cell the
monopoles have more zero modes due to the Kaluza-Klein modes,
suggesting that it is not fundamental. Indeed, it is a  composite of
instantons and fundamental monopoles as shown in
Ref.~\cite{piljin}. Depending on the gauge group, the fundamental cell may
not be the smallest cell for which all of $x_4$-independent $\bbeta_i$
monopoles have only four zero modes.

For the $x_4$-dependent $\bbeta_0$ monopole solutions, let us proceed
first with the (\ref{time1}) solution. For these solutions we can
gauge away the $x_4$-dependence by a large gauge transformation, and
then analysis is identical as before. The fluctuation should be
single-valued in $x_4$ and one use the parameter $y$ of
Eq.~(\ref{ypara}) with ${\bf h}_{\rm new}$  of Eq.~(\ref{hnew}).
The number of these fundamental monopoles still have four zero modes
in the fundamental cell.

Instead we could use the solution (\ref{time2}), which is also only option
for the $G_2,F_4,E_8$ cases. To do fluctuation analysis around this
solution, we notice that the fluctuation should be still
single-valued. Again the fluctuation equation (\ref{fluct}) can be
split into the irreducible representations with respect to ${\bf
t}(\bbeta_0)$. Only $t=1$ fluctuations are along $\bbeta_0$ itself,
since $\bbeta_0 $ is one of the longest.  Any other nontrivial
fluctuations have $t=\frac{1}{2}$.  {}From the flutuation
equation~(\ref{fluct}), we can gauge away $e^{-ix_4 t^3(\beta_0)}$
from the background field, as  change of variables, which makes the $t=1$
fluctuations still single-valued but the $t=1/2$ fluctuations
double-valued. The $y$-parameter is then 
\beq
y= \frac{{\bf h}\cdot\balpha}{m_0} -\bbeta_0^*\cdot \balpha +
\frac{\bbeta^*\cdot \balpha + l}{m_0}.
\eeq
Thus the $l$ parameter in Eq.~(\ref{ypara}) would be integer for $t=1$
fluctuations and half-integer for $t=1/2$ fluctuations. Of course this
analysis in theories with group of nontrivial center is consistent
with the analysis done before by using a large gauge transformation.
For all gauge groups including $G_2,F_4,E_8$, one can check after some
labor that a single $\bbeta_0$ monopole described by Eq.~(\ref{time2})
has indeed only four zero modes if ${\bf h}$ lies in the fundamental
cell. Also one can show that the fundamental cell is the maximal
region after some labor.

Thus there  exist $r+1$ distinct self-dual magnetic monopoles, one for
each root in the extended Dynkin diagram. For ${\bf h}$ in the
fundamental cell, each of these monopole solutions have only four zero
modes. In addition, the  topological charge $m_i, 0\le i\le r$ are
positive in the fundamental cell. Their magnetic
charge is $4\pi \bbeta_i^*$. Since the instanton carries zero magnetic
charge, one can ask whether there is a linear combination of these
magnetic charges such that the total magnetic charge is zero. 
It turns out that there is a unique minimum set of positive integers
$(n_0,n_1,...,n_r)$ as shown in Table II,  such that
\beq
\sum_{i=0}^r n_i \bbeta_i^* = 0 .
\label{dynkin}
\eeq
Especially the number of $\bbeta_0$ monopoles  is  $n_0=1$. 
We call these positive integers to be Dynkin numbers, whose sum turns
out to be the dual Coxeter number
\beq
c_2(G) = \sum_{i=0}^r n_i .
\eeq
Let us now imagine nonlinearly superposed field configurations for  
fundamental monopoles whose numbers are prescribed by  Dynkin
numbers. Then their  net magnetic charge is equal to zero and their total
topological charge is 
\beq
\sum_{i=0}^r n_i m_i = 1 + {\bf h}\cdot \sum_{i=0}^r n_i \bbeta_i^* = 
1.
\eeq
The number of zero modes is $4c_r(G)$. They satisfy the self-dual
equations.  This is exactly what one expects for a single instanton
solution.

{} 

\vspace{4mm}
\begin{center}
\begin{small}
\begin{tabular}{|l|l|l|l|l|}\hline
Group  & $r$ &  $d(G)$  & $c_2(G)$ &  $(n_0,n_1,...,n_r)$  \\ \hline
$SU_{n+1}$ & $n$ & $n^2+n$ & $n+1$ &  $(1,1,...,1)$ \\ \hline
$SO_{2n+1}$ {} & $n$ & $n(2n+1)$ & $2n-2$   & $(1,1,1,2,2,...,2)$ \\
\hline
$SO_{2n}$ & $n$ & $n(2n-1)$ & $2n-2$  & $(1,1,1,1,2,2,...,2)$
\\ \hline
$Sp_{2n}$ & $n$ & $n(2n+1)$ & $n+1$ & $(1,1,...,1)$ \\
\hline
$G_2$ & $2$ & $14$ & $4$ & $(1,1,2)$ \\ \hline
$F_4$ & $4$ & $52$ & $9$ & $(1,1,2,2,3)$ \\ \hline
$E_6$ & $6$ & $78$ & $12$ & $(1,1,1,2,2,2,3)$ \\ \hline
$E_7$ & $7$ & $133$ & $18$    & $(1,1,2,2,2,3,3,4)$ \\ \hline
$E_8$ & $8$ & $248$ & $30$   & $(1,2,2,3,3,4,4,5,6)$ \\ \hline 
\end{tabular}
\end{small} \\
\vspace{4mm} {\bf Table II}:  The rank $r$, dimension $d(G)$, the dual
Coxeter number $c_2(G)$, and Dynkin numbers, for all simple compact Lie
groups.  
\end{center}

To conclude, we have shown that there exist $r+1$ distinct self-dual
magnetic monopoles on $R^3\times S^1$ and that instantons on
$R^3\times S^1$ are made of these magnetic monopoles. Quite reasonable
is our assumption that all self-dual solutions with the same boundary
condition can be put together.  It would be interesting to see whether
this relation between instantons and magnetic monopoles can be
explored further in the various area mentioned at the beginning: the
finite temperature QCD~\cite{lattice}, the generalization of
Witten-Seiberg results of $N=2$ or $N=1$ supersymmetric theories on
$R^3\times S^1$~\cite{seiberg,cumrun}, and D-brane and
M-theory~\cite{piljin}. Especially recent works by Vafa
et.al.~\cite{cumrun} on $N=1$ supersymmetric theories on $R^3\times
S^1$ contains the dual Coxeter numbers and Dynkin numbers by using $F$
and $M$ theories, which we believe can be interpreted as the effect
due to the presence of constituent magnetic monopoles for instantons.

\vspace{4mm}

\centerline{\bf Acknowledgments} 

This work is supported in part by the U.S. Department of Energy. We
thank Erick Weinberg for useful discussions.

\end{document}